\documentclass{article}
\usepackage{dcase2018,amsmath,graphicx,url,times,booktabs,tabularx}

\usepackage{multirow}
\usepackage{hyperref}
\usepackage{stmaryrd}
\usepackage{stfloats}
\usepackage{amssymb}

\title{COMBINING HIGH-LEVEL FEATURES OF RAW AUDIO WAVES AND MEL-SPECTROGRAMS FOR AUDIO TAGGING}

\twoauthors{Marcel Lederle\sthanks{Both authors contributed equally to this work.}}
    {University of Konstanz\\
     Konstanz, Germany\\
     marcel.lederle@uni.kn}
  {Benjamin Wilhelm\footnotemark[1]}
    {University of Konstanz \\
     Konstanz, Germany\\
     benjamin.wilhelm@uni.kn}

\begin{document}

\ninept
\maketitle

\begin{sloppy}
\begin{abstract}
In this paper, we describe our contribution to Task 2 of the DCASE 2018 Audio Challenge~\cite{fonseca2018general}. While it has become ubiquitous to utilize an ensemble of machine learning methods for classification tasks to obtain better predictive performance, the majority of ensemble methods combine predictions rather than learned features. We propose a single-model method that combines learned high-level features computed from log-scaled mel-spectrograms and raw audio data. These features are learned separately by two Convolutional Neural Networks, one for each input type, and then combined by densely connected layers within a single network. This relatively simple approach along with data augmentation ranks among the best two percent in the Freesound General-Purpose Audio Tagging Challenge on Kaggle.

\end{abstract}


\section{INTRODUCTION}\label{sec:intro}
For humans, it seems to be effortless to associate sounds with events or categories that describe the perceived sound best. However, the complex structure and the large amount of information transmitted through sound makes it particularly difficult to extract that information automatically.

Recognizing a wide variety of sounds has many applications in our today's life. These include surveillance~\cite{harma2005automatic, crocco2016audio, valenzise2007scream}, acoustic monitoring ~\cite{goetze2012acoustic}, and automatic description of multimedia~\cite{jin2016video}. Due to the diversity of sounds belonging to the same category, a reliable recognition of manifold sound categories is still under ongoing research.

Carefully hand-crafted features such as Mel Frequency Cepstral Coefficients (MFCCs) were the dominant features used for speech recognition~\cite{ittichaichareon2012speech, logan2000mel} and music information retrieval~\cite{tzanetakis2002musical}, but the trend is now shifting toward deep learning~\cite{abdel2013exploring}. The approach of manual feature engineering has drawbacks compared to deep learning based methods because it requires considerable effort and expertise to manually create features for a specific purpose. In particular, most of the engineered features, such as MFCCs and spectral centroids~\cite{eronen2006audio}, are non-task specific, whereas all deep learning approaches are task-specific due to its formulation as a minimization process on task-specific training examples. Since Convolutional Neural Networks (CNNs) have shown a remarkable progress in visual recognition tasks~\cite{russakovsky2015imagenet} over the last years, it has become common to use CNNs for feature extraction and classification in the audio domain~\cite{abdel2014convolutional, piczak2015environmental}.
Several CNN architectures, such as AlexNet~\cite{krizhevsky2012imagenet}, VGG~\cite{simonyan2014very}, ResNet~\cite{he2016deep}, and Inception-v3~\cite{szegedy2016rethinking}, have been proposed for image classification, which are also well suited to the task of audio tagging.

The goal of Task 2 of the DCASE 2018 Challenge~\cite{fonseca2018general} was to predict the category of an audio clip belonging to one out of 41 heterogeneous classes, such as ``Acoustic guitar'', ``Bark'', ``Bus'' and ``Telephone'', drawn from the AudioSet Ontology~\cite{gemmeke2017audio}. Training and testing data contain a diverse set of user-generated audio clips from Freesound (\href{https://freesound.org}{https://freesound.org})~\cite{fonseca2018general}.

In this paper, we mainly focus on building an audio-tagging system that uses both the raw audio data and the corresponding mel-spectrogram rather than ensembling~\cite{breiman1996bagging} or stacking ~\cite{dvzeroski2004combining} multiple classifiers. 

%

\section{METHOD}\label{sec:method}
Our audio-tagging system comprises two separately trained Convolutional Neural Networks on raw audio and mel-spectrogram, respectively. The learned high-level features are then combined by a densely connected neural network to form the system. In the following, we describe each model in detail.


\subsection{CNN on Raw Audio (\textit{cnn-audio})}\label{ssec:cnnaudio}

For the \textit{cnn-audio} model, we use an architecture similar to common architectures for image classification, like VGG16~\cite{simonyan2014very} or AlexNet~\cite{krizhevsky2012imagenet}, but with one-dimensional convolutions and one-dimensional max pooling. As described in Table~\ref{tab:cnnaudio}, we use four blocks, each consisting of two convolutional and one max-pooling layer. The number of filters is increased in each consecutive block, while the kernel size is decreased. The pool-size of the max-pooling layers is chosen to quickly reduce the large time dimension.
After each block and after the dense layer, we apply batch normalization~\cite{ioffe2015batch}, as experiments have shown that it reduces the training time and increases the model accuracy.
To introduce nonlinearities into our network, we apply a ReLU activation function~\cite{nair2010rectified} after each convolutional layer and dense layer.


\begin{table}[t]
\centering
\begin{tabularx}{\columnwidth}{lccc}
\toprule
Layer            & 1sec shape  & 2sec shape  & 3sec shape   \\ \midrule \midrule
input            & (44100, 1)  & (88200, 1)  & (132300, 1)  \\ \midrule
conv1d, 11, 32   & (44100, 32) & (88200, 32) & (132300, 32) \\
conv1d, 11, 32   & (44100, 32) & (88200, 32) & (132300, 32) \\
max-pool1d, 8/16 & (5512, 32)  & (5512, 32)  & (8268, 32)   \\ \midrule
conv1d, 9, 64    & (5512, 64)  & (5512, 64)  & (8268, 64)   \\
conv1d, 9, 64    & (5512, 64)  & (5512, 64)  & (8268, 64)   \\
max-pool1d, 16   & (344, 64)   & (344, 64)   & (516, 64)    \\ \midrule
conv1d, 7, 128   & (344, 128)  & (344, 128)  & (516, 128)   \\
conv1d, 7, 128   & (344, 128)  & (344, 128)  & (516, 128)   \\
max-pool1d, 16   & (21, 128)   & (21, 128)   & (32, 128)    \\ \midrule
conv1d, 5, 256   & (21, 256)   & (21, 256)   & (32, 256)    \\
conv1d, 5, 256   & (21, 256)   & (21, 256)   & (32, 256)    \\
max-pool1d, 16   & (1, 256)    & (1, 256)    & (2, 256)     \\ \midrule
dense, 512       & (512)       & (512)       & (512)        \\
softmax, 41      & (41)        & (41)        & (41)         \\ \bottomrule
\end{tabularx}
\caption{The architecture of the \textit{cnn-audio} model. Note that for the one-second model, the first max-pooling layer uses a pool-size of 8, while for other models a pool-size of 16 is used.}\label{tab:cnnaudio}
\end{table}

\subsection{CNN on Mel-Spectrogram (\textit{cnn-spec})}\label{ssec:cnnspec}

The \textit{cnn-spec} model is a two-dimensional convolutional neural network taking mel-spectrograms as input. The architecture is again similar to common image classification architectures and is described in detail in Table~\ref{tab:cnnspec}. As with the one-dimensional model, we apply batch normalization after each block and the dense layer and use the ReLU activation function after convolutional and dense layers.

The mel-spectrogram is extracted using librosa~\cite{brian_mcfee_2018_1174893} with the original sampling frequency of $44.1$ kHz, 2048 FFT points, 128 mel-bins, and a hop-length of 256. The amplitude of the mel-spectrogram is scaled logarithmically, and the scaled mel-spectrogram is resized in time dimension to fit the model input size.

\begin{table}[t]
\centering
\setlength{\tabcolsep}{1pt}
\begin{tabularx}{\columnwidth}{@{}lccc@{}}
\toprule
Layer                      & 1sec shape     & 2sec shape     & 3sec shape     \\ \midrule \midrule
input                      & (128, 170, 1)  & (128, 300, 1)  & (128, 400, 1)  \\ \midrule
conv2d, 4$\times$4, 64     & (128, 170, 64) & (128, 300, 64) & (128, 400, 64) \\
conv2d, 4$\times$4, 64     & (128, 170, 64) & (128, 300, 64) & (128, 400, 64) \\
max-pool2d, 1$\times$1/2$\times$2 & (128, 170, 64) & (64, 150, 64)  & (64, 200, 64) \\ \midrule
conv2d, 4$\times$4, 64     & (128, 170, 64) & (64, 150, 64) & (64, 200, 64)   \\
max-pool2d, 2$\times$2     & (64, 85, 64)   & (32, 75, 64)   & (32, 100, 64)  \\ \midrule
conv2d, 3$\times$3, 128    & (64, 85, 128)  & (32, 75, 128)  & (32, 100, 128) \\
max-pool2d, 2$\times$4     & (32, 21, 128)  & (16, 18, 128)  & (16, 25, 128)  \\ \midrule
conv2d, 3$\times$3, 128    & (32, 21, 128)  & (16, 18, 128)  & (16, 25, 128)  \\
max-pool2d, 2$\times$2     & (16, 10, 128)  & (8, 9, 128)    & (8, 12, 128)   \\ \midrule
conv2d, 3$\times$3, 256    & (16, 10, 256)  & (8, 9, 256)    & (8, 12, 256)   \\
max-pool2d, 2$\times$2     & (8, 5, 256)    & (4, 4, 256)    & (4, 6, 256)    \\ \midrule
conv2d, 3$\times$3, 256    & (8, 5, 256)    & (4, 4, 256)    & (4, 6, 256)    \\
max-pool2d, 2$\times$2     & (4, 2, 256)    & (2, 2, 256)    & (2, 3, 256)    \\ \midrule
dense, 256                 & (256)          & (256)          & (256)          \\
softmax, 41                & (41)           & (41)           & (41)           \\ \bottomrule
\end{tabularx}
\caption{The architecture of the \textit{cnn-spec} model. Note that the first max-pooling layer does not exist in the case of the one-second model.}\label{tab:cnnspec}
\setlength{\tabcolsep}{6pt}
\end{table}


\subsection{Joining CNNs (\textit{cnn-comb})}\label{ssec:stacking}
We remove the softmax and dense layer of both the trained \textit{cnn-audio} and \textit{cnn-spec} model and then concatenate the output features of the previous layer of both models so as to join them.
The concatenated features are then connected to a densely connected neural network with four hidden layers. The hidden dense layers have 512, 256, 256, and 128 neurons, respectively. The complete model is illustrated in Figure~\ref{fig:architecture}.

We train \textit{cnn-audio} and \textit{cnn-spec} from scratch. Afterward, the weights of these models are transferred to the \textit{cnn-comb} model and only the newly added dense layers are trained. The splitting of the training of \textit{cnn-comb} into three steps, facilitates the procedure.

\begin{figure*}[h]
  \centering
  \centerline{\includegraphics[width=\textwidth]{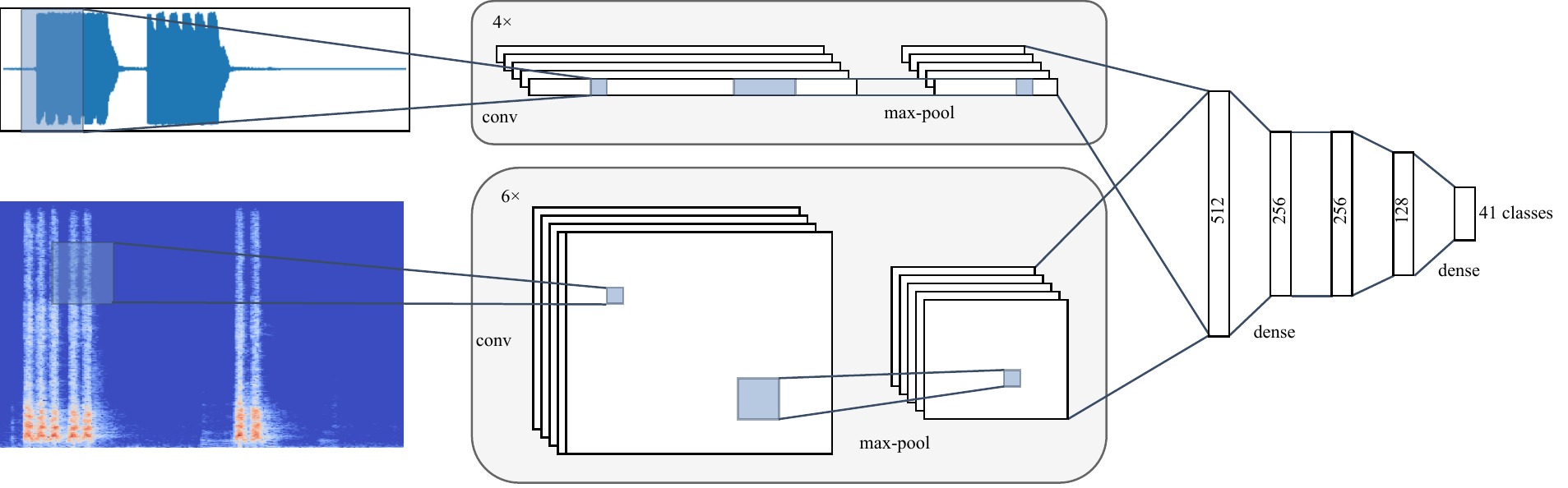}}
  \caption{Illustrated architecture of the complete model.}\label{fig:architecture}
\end{figure*}

\subsection{Data Augmentation}\label{ssec:dataaug}
To prevent our model from overfitting, we make use of extensive data augmentation during training (time shifting, cropping, padding, and blending clips of same and different categories). Each of these augmentation techniques is applied to the raw audio wave and the mel-spectrogram. In the remaining section, we explain the augmentation methods based on the raw audio wave.

First, we apply a uniformly random time shift to the audio clip. To ensure that the audio clips fit the size of the model input, crops are taken from too long audio files and too short audio files are padded.
For audio clips that are longer than the model input size, we use a crop with the size of the model input taken from a random position.
If an audio sample fits multiple times ($n$ times) in the input size, it is replicated such that it appears $k \in \{1, \ldots, n \}$ times with a probability of $1/n$, and the remaining space before, after, and in between replications is filled with zeros.

Additionally, we enhance the robustness of our model by blending multiple audio clips of same or different categories. This method is referred to as mixup~\cite{zhang2018mixup}. We blend them by assigning a random weight to each sample (weights sum up to one) and taking the weighted sum. If the blended samples are of the same class, the model should still predict the common class for the newly generated training sample. In the other case, if the blended samples are of distinct classes, the model is trained to predict the weight of each included class.

While the spectrogram is computed in advance, the computationally inexpensive data augmentation techniques can be computed on-the-fly during training. This saves disk-space and guarantees a large amount of diverse training data.

\subsection{Implementation Details}\label{ssec:impl}

We implemented the described method using Keras~\cite{chollet2015keras} in Python.

To monitor overfitting and the model performance during training, we exclude a part of the training data as validation data. To still make use of all training data, we train five models on stratified folds of the training data such that each training example is used once for validation and four times for training. For the final prediction, we accumulate the predictions of all five models using the geometric mean.

All models are trained using the Adam optimizer~\cite{kingma2014adam} with a fixed learning rate of $0.001$ and a mini-batch size of $32$ for a maximum of $300$ epochs, but stopping earlier if the validation loss hasn't improved for $35$ epochs. We use the categorical cross-entropy loss function and weight the loss according to the distribution of training examples per class, thereby ensuring that the models pay more attention to samples from an under-represented class.

Because many clips contain silence, we cut off silent parts at the beginning and at the end of an audio clip that do not exceed a volume of $40$ decibel.

When predicting the test data, we have to take the varying length of audio files into account.
It is not sufficient to only predict on one crop of too long tracks because important features might not be present in the selected crop. Therefore, we run the inference on many crops of the audio file with a step size of $5120$ frames, which is approximately $0.12$ seconds.
For too short audio tracks, the model might be able to better recognize class-specific features in certain parts of the input. Therefore, we generate multiple inputs by padding the audio file with zeros such that the real audio appears at different positions in the input. Again, we use a step size of $5120$ frames.
 Multiple predictions for one audio file are combined by means of the geometric mean.

\section{EVALUATION}\label{sec:evaluation}

\subsection{Dataset}

We evaluate our method on the dataset provided for Task 2 of the DCASE 2018 Challenge~\cite{fonseca2018general}, which comprises $9473$ training and $1600$ test samples. The test data set has been manually verified, whereas the training data features labels of different reliability. Each mono audio file has a bit-depth of $16$, a sampling rate of $44.1$ kHz, and is associated with one out of $41$ classes of the AudioSet Ontology~\cite{gemmeke2017audio}.
  
The class distribution of both training and test set is not balanced and ranges from $94$ to $300$ and from $25$ to $110$ samples per class, respectively. The duration of the shortest audio file is $300$ ms and $30$ seconds for the longest clip, while the average length is $6.8$ seconds for the train set and $5.2$ seconds for the test set. 

\subsection{Metric}

The Challenge uses mean Average Precision at three (mAP@3) for evaluating test results, which allows up to three predictions per audio clip. Full credit is given if the first prediction matches the label of the clip, while less credit is given if one of the other predictions is correct. 
The evaluation metric is defined as \[\text{mAP@3} = \frac{1}{U} \sum_{i=1}^{U} \sum_{j=1}^{\min(3,n_i)} \frac{\llbracket y_{ij} = \hat{y}_i\rrbracket}{j},\] where $U$ is the total number of scored audio files, $y_{ij}$ is the predicted label for file $i$ at position $j$, $\hat{y}_i$ is the ground-truth label for file $i$, $n_i$ is the total number of predicted labels for file $i$, $\llbracket \texttt{True} \rrbracket = 1, \llbracket \texttt{False} \rrbracket = 0$. No label may be predicted multiple times for one audio file.

\subsection{Results}

We have trained all models on inputs of one, two, and three seconds, as described in Section~\ref{ssec:impl}, and evaluated the model performance on the test set (see Table~\ref{tab:results}). We have observed that for each crop length, the combined model performs significantly better compared to models with a single input. Combined models with an input size of two and three seconds, perform best and rank in the upper two percent on the Private Leaderboard on Kaggle.

\begin{table}[t]
\centering
\begin{tabularx}{\columnwidth}{XXXXX}
\toprule
Model                       & Crop length & Public score (mAP@3) & Private score (mAP@3) & Total (mAP@3) \\
\midrule \midrule
\multirow{3}{*}{\textit{cnn-audio}} & 1sec & 0.920 & 0.888 & 0.894 \\
				    & 2sec & 0.921 & 0.884 & 0.891 \\
				    & 3sec & 0.935 & 0.889 & 0.898 \\
\midrule
\multirow{3}{*}{\textit{cnn-spec}}  & 1sec & 0.930 & 0.923 & 0.924 \\
				    & 2sec & 0.950 & 0.928 & 0.932 \\
				    & 3sec & 0.935 & 0.930 & 0.931 \\
\midrule
\multirow{3}{*}{\textit{cnn-comb}}  & 1sec & 0.955 & 0.939 & 0.942 \\
				    & 2sec & \textbf{0.966} & \textbf{0.944} & \textbf{0.948} \\
				    & 3sec & 0.956 & \textbf{0.944} & 0.946\\
\bottomrule
\end{tabularx}
\caption{Evaluation results of the individual models on the public ($301$ samples), private ($1299$ samples), and full test set.}\label{tab:results}
\end{table}

Additionally, we determined the per-category mAP@3 on the complete test set showing that some classes are more challenging to predict than others (see Table~\ref{tab:perclassmap}). Our model primarily struggles with the classes ``Squeak'', ``Telephone'' and ``Fireworks'', but it still beats the baseline system~\cite{fonseca2018general} in every per-category mAP@3 score.

\begin{table*}[t]
\centering
\setlength{\tabcolsep}{3pt}
\begin{tabularx}{\textwidth}{@{}Xccc|Xccc|Xccc@{}}
\toprule
\textbf{Name} & \textbf{samples} & \textbf{time} & \textbf{mAP@3} &
\textbf{Name} & \textbf{samples} & \textbf{time} & \textbf{mAP@3} &
\textbf{Name} & \textbf{samples} & \textbf{time} & \textbf{mAP@3} \\
\midrule
Acoustic\_guitar        &      300 &  52.2 & 0.893 &
Electric\_piano         &      150 &  25.5 & 1.000 &
Microwave\_oven         &      146 &  25.1 & 0.966 \\
Applause                &      300 &  58.2 & 1.000 &
Fart                    &      300 &  18.6 & 0.944 &
Oboe                    &      299 &  15.3 & 0.976 \\
Bark                    &      239 &  44.6 & 0.982 &
Finger\_snapping        &      117 &   5.9 & 1.000 &
Saxophone               &      300 &  33.7 & 0.942 \\
Bass\_drum              &      300 &  12.8 & 1.000 &
Fireworks               &      300 &  48.2 & 0.786 &
Scissors                &       95 &  15.7 & 0.927 \\
Burping\_or\_eructation &      210 &  11.7 & 1.000 &
Flute                   &      300 &  46.2 & 1.000 &
Shatter                 &      300 &  26.1 & 0.960 \\
Bus                     &      109 &  28.4 & 0.953 &
Glockenspiel            &       94 &   8.4 & 0.856 &
Snare\_drum             &      300 &  17.9 & 0.912 \\
Cello                   &      300 &  37.3 & 0.951 &
Gong                    &      292 &  41.8 & 0.968 &
Squeak                  &      300 &  38.2 & 0.603 \\
Chime                   &      115 &  23.8 & 0.891 &
Gunshot\_or\_gunfire    &      147 &  11.1 & 0.950 &
Tambourine              &      221 &  10.1 & 0.975 \\
Clarinet                &      300 &  34.7 & 0.991 &
Harmonica               &      165 &  18.6 & 0.970 &
Tearing                 &      300 &  38.7 & 0.981 \\
Computer\_keyboard      &      119 &  23.0 & 1.000 &
Hi-hat                  &      300 &  18.6 & 0.957 &
Telephone               &      120 &  16.2 & 0.788 \\
Cough                   &      243 &  22.4 & 1.000 &
Keys\_jangling          &      139 &  18.8 & 0.929 &
Trumpet                 &      300 &  28.3 & 0.959 \\
Cowbell                 &      191 &  10.9 & 1.000 &
Knock                   &      279 &  19.6 & 0.957 &
Violin\_or\_fiddle      &      300 &  26.6 & 0.986 \\
Double\_bass            &      300 &  16.9 & 0.946 &
Laughter                &      300 &  36.3 & 0.974 &
Writing                 &      270 &  48.3 & 0.948 \\
Drawer\_open\_or\_close &      158 &  18.0 & 0.925 &
Meow                    &      155 &  18.7 & 1.000 \\
\bottomrule
\end{tabularx}

\caption{Per category mAP@3 score of the \textit{cnn-comb} 2sec model on the full test set and the number of samples along with time in minutes of the respective class in the train set.}\label{tab:perclassmap}
\setlength{\tabcolsep}{6pt}
\end{table*}

To verify that the performance gain of the combined model results from combining the extracted high-level features from both models, we compare \textit{cnn-audio} and \textit{cnn-spec} to the \textit{cnn-comb} model, where one of its inputs is set to zero (See Figure~\ref{fig:audio-spec-comparison}).

For categories in which the single \textit{cnn-audio} model outperforms the single \textit{cnn-spec} model, the combined model performs better if the audio input is present and not set to zero. Otherwise, if \textit{cnn-spec} outperforms \textit{cnn-audio}, \textit{cnn-comb} has a higher score if the mel-spectrogram is present and not set to zero. Setting either the mel-spectrogram or the audio wave to zero forces the \textit{cnn-comb} model to make predictions based on a single input. 

\textit{cnn-comb} performs equally or better than \textit{cnn-comb} with one single input set to zero because 
it has learned to utilize meaningful high-level features of both inputs jointly, which are not given by zeroed inputs. For the same reason, \textit{ccn-comb} with one zeroed input usually performs worse than the corresponding model with a single input.


We conclude that \textit{cnn-comb} makes use of the learned high-level features from both \textit{cnn-audio} and \textit{cnn-spec}, but it focuses more on features belonging to the superior single model.

\begin{figure*}[t]
\centering
	\includegraphics[width=\textwidth]{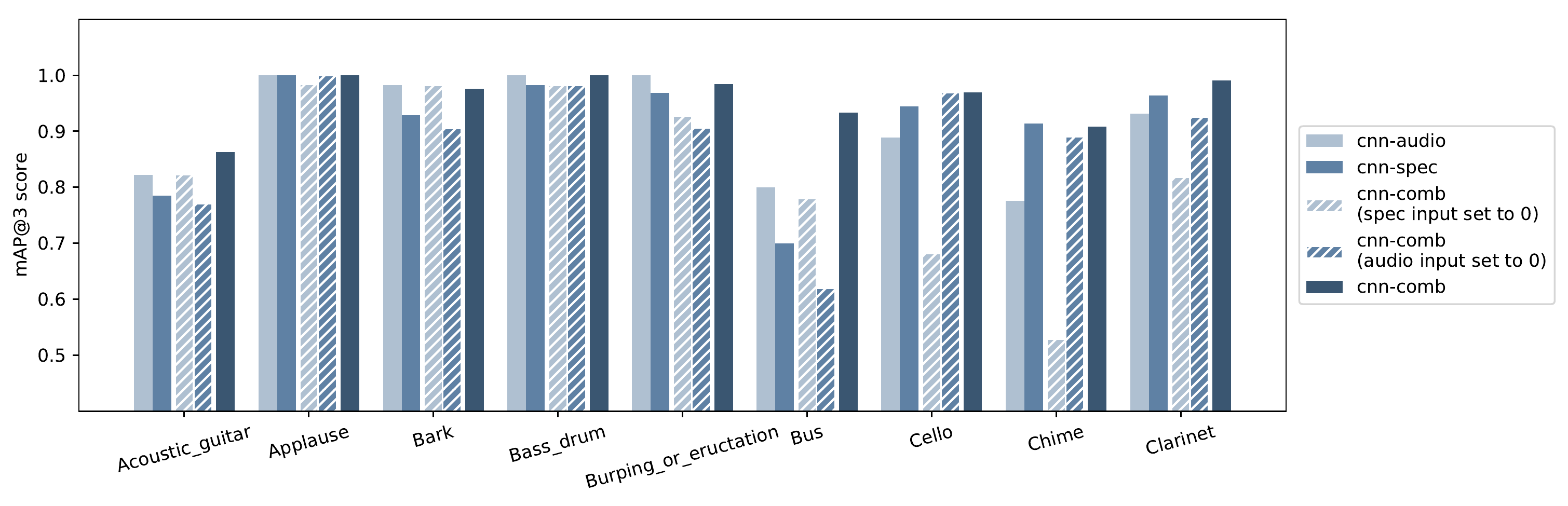}
	\caption{Comparison of per-category scores of single input models, combined models with one input alternately set to zero, and the combined model with both inputs. The mAP@3 score is reported on a single fold for each model.}\label{fig:audio-spec-comparison}
\end{figure*}

\section{CONCLUSION}\label{sec:conclusion}

In this paper, we have proposed a method for audio-tagging that extends current Convolutional Neural Network approaches that only make use of a frequency representation by adding a second input that incorporates the raw audio wave. Adding the additional input, has improved the mAP@3 score significantly. We have demonstrated the capabilities of our model by competing in the Freesound General-Purpose Audio Tagging Challenge on Kaggle and ranking in the top two percent of all participants.


\section{ACKNOWLEDGMENT}\label{sec:ack}
We thank Christian Borgelt and Christoph Doell for motivating us to take part in the Kaggle competition and Christoph Doell for his valuable comments on the manuscript.

\bibliographystyle{ieeetran}

\end{sloppy}
\end{document}